\documentclass{elsart} 
\usepackage{epsfig} 
\usepackage{bm} 
\usepackage{color}

\renewcommand{\it}[1]{\textit{#1}} 
\newcommand{\Onecol} 
{\begin{widetext} 
\onecolumngrid} 
\newcommand{\Twocol} 
{\end{widetext} \twocolumngrid}
 
\newcommand{\be}{\begin{equation}} 
\newcommand{\ba}{\begin{array}} 
\newcommand{\bea}{\begin{eqnarray}} 
\newcommand{\bfi}{\begin{figure}} 
\newcommand{\ee}{\end{equation}} 
\newcommand{\ea}{\end{array}} 
\newcommand{\eea}{\end{eqnarray}} 
\newcommand{\efi}{\end{figure}}

\begin{document}
\begin{frontmatter}
\title{Excitable Dynamics in the Presence of Time Delay}  
\author{Gautam C Sethia$^{1}$ and Abhijit Sen$^{2}$}
\address{Institute for Plasma Research, Bhat, Gandhinagar 382 428, India}
\thanks{E-mail: gautam@ipr.res.in\\ $^2$ E-mail: abhijit@ipr.res.in}
\begin{abstract}
 The spiking properties of a subcritical Hopf oscillator with a time delayed nonlinear feedback is investigated. Finite time delay is found to significantly affect both the statistics and the fine structure of the spiking behavior.  These dynamical changes are explained in terms of the fundamental modifications occurring in the bifurcation scenario of the system. Our mathematical
model can find useful applications in understanding the dynamical behaviour of various real life excitable systems where propagation delay effects are ubiquitous.
\end{abstract}
\begin{keyword}
Excitability \sep spiking \sep Hopf oscillator  \sep time delayed feedback 
\PACS 05.45.-a \sep 02.30.Ks \sep 02.50.Ey \sep 05.40.Ca
\end{keyword}
\end{frontmatter}
\section{INTRODUCTION}

Excitability, which is the property responsible for the generation of spiky or bursty output in dynamical systems, continues to be an active area of investigation in various branches of physics, chemistry, biology and nonlinear sciences \cite{linder,arkady}. The spiky action potentials of neurons, for example, are the subject of interest for neurophysiologists as well as for researchers interested in the mathematical modeling of the brain through the study of neural networks \cite{frank}. From a mathematical viewpoint a system is said to be excitable if a small perturbation away from its stable equilibrium  state can result in a large excursion of its potential (known variously as spiking or firing) before returning to its quiescence. These large excursions exist because the system equilibrium is close to bifurcations from rest to oscillatory states and the nature of excitability is intimately related to the type of bifurcation \cite{izh2000}. 

Various existent mathematical models provide
good representations  of the qualitative features of system excitability. The most well known and extensively studied models are those associated with the behaviour of individual or ensembles of neurons \cite{hodgkin}. An issue that has received only limited attention  in these studies is the effect of time delays \cite{foss} on the nature of the bifurcation behavior and the concommitant effect on excitability. Time delays, arising from finite conduction/propagation speeds, can certainly be significant in situations where neurons
(or individual cells of a network)  communicate over some distance, and can influence, for example, the synchronization condition for an ensemble of cells. The question we ask in this paper is whether even in a single neuron there can exist time delay induced changes in its intrinsic excitability properties. To address this issue we introduce and investigate a simple mathematical model consisting of a subcritical Hopf oscillator with a time delayed nonlinear feedback. Our model contains the necessary ingredients to reproduce most of the basic features of excitability that are displayed by standard neuronal models such as the Hodgkin-Huxley system \cite{izh2000} and in addition provides a convenient means of studying the effect of time delay on its neural dynamical behavior. Subcritical Hopf bifurcation systems with delayed feedback have been studied in the past using different model equations \cite{larger} and in the context of applications like control of unstable periodic orbits \cite{pyragas}. To the best of our knowledge the effect of time delay on the spiking properties of a single neuron model has not been investigated so far. Our basic idea of incorporating a nonlinear feedback term has been motivated to some extent by the structure of the recently discovered autapse neurons which have axons terminating back on their own dendrites \cite{tamas,chr2004}. However the model can have wider applicability. 

In the absence of time delay our model system shows a \textit{saddle-node on a limit cycle} (SNLC) bifurcation 
for certain strengths of the nonlinear feedback and when subjected to an external periodic stimulus and noise exhibits typical neural excitability features in the form of integer multiple spiking (IMS) \cite{longtin}. We find that the excitability properties can be significantly altered in the presence of finite time delay which affects both the statistics of spiking behavior as well as the fine structure of the individual spikes. We analyze these changes through a study of the multimodal histogram of interspike intervals (ISI) and other characteristic signatures. We also carry out a detailed bifurcation analysis and find that in the presence of time delay the original bifurcation gives way to a saddle separatrix loop (SSL) bifurcation beyond a certain threshold value of the time delay. The resultant changes in the spiking behavior are explained in terms of this topological modification.\\

\begin{figure}
\centerline{\includegraphics[width=12.0cm,height=8cm]{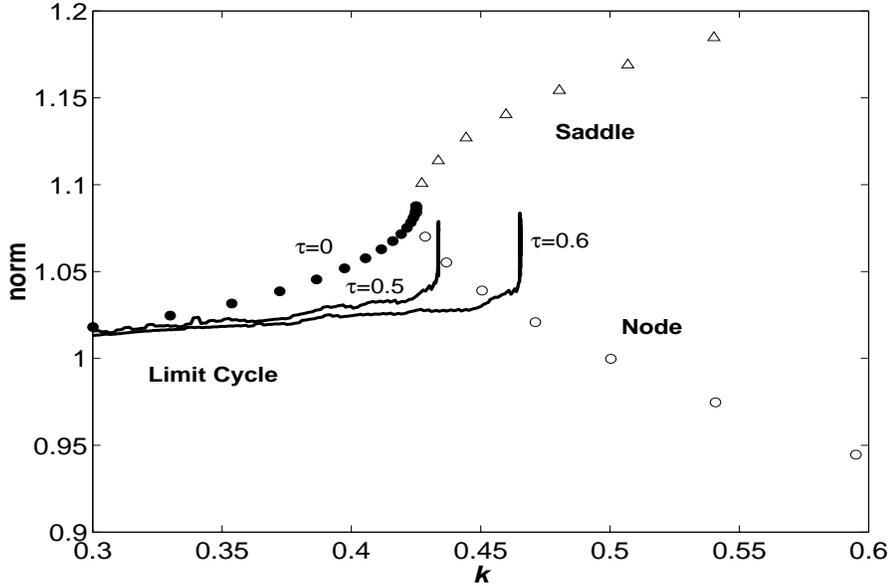}}
\caption{Bifurcation diagrams as a function of the feedback strength $k$ with and without time delay.}
\label{bif_k}
\end{figure}

\begin{figure}
\centerline{\includegraphics[width=12.0cm,height=8cm]{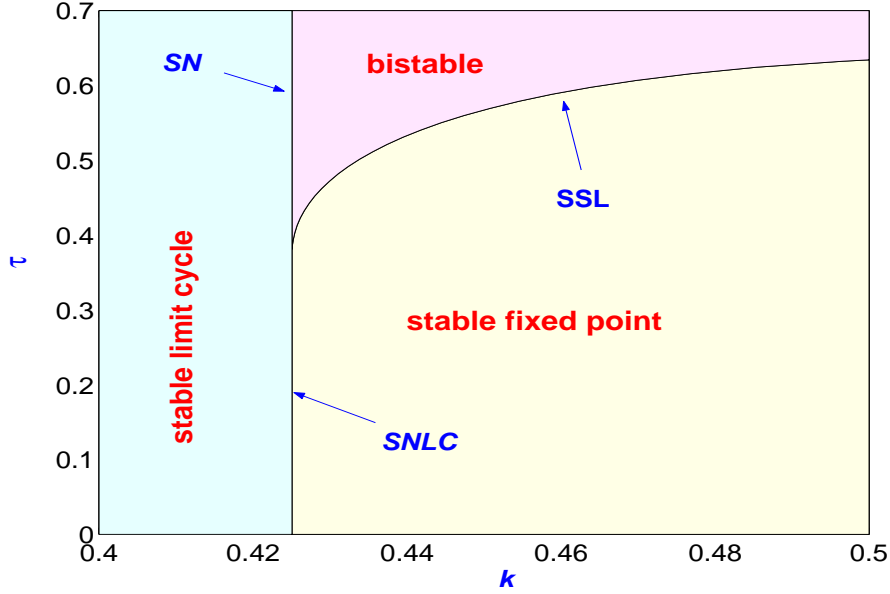}}
\caption{ (color online) Stability diagram in the parameter space of $k$ and $\tau$. Note the various bifurcation boundaries demarcating the different stability regions.}
\label{tau_k}
\end{figure}

\begin{figure}
\centerline{\includegraphics[width=12cm,height=8cm]{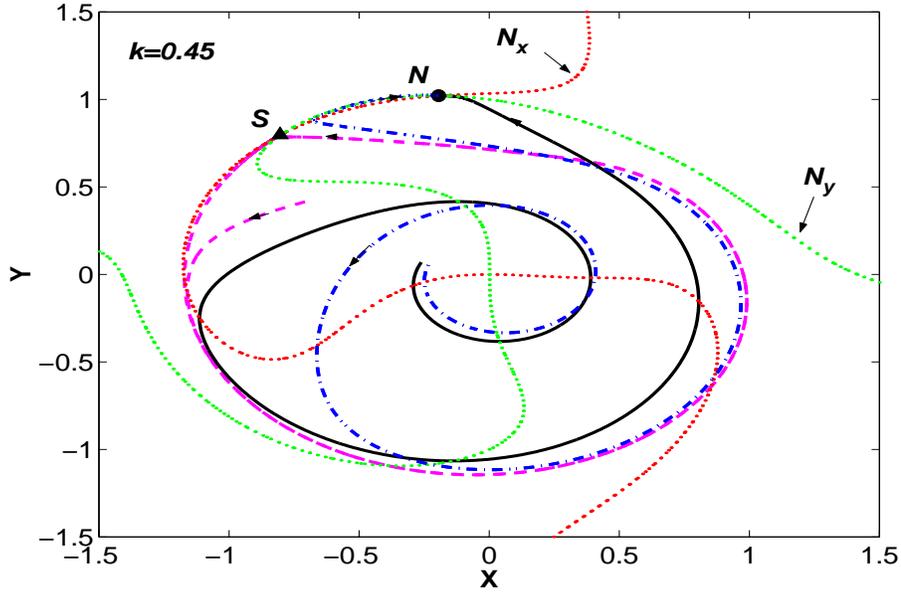}}
\caption{(color online) Typical trajectories for various values of $\tau$ ($\tau =0:$ solid; $\tau=0.5:$ dash-dot; $\tau=0.57:$ dash) illustrating an SSL bifurcation beyond a threshold $\tau_c$. The nullclines (\textit{$N_{x}$,$N_{y}$}) along with the saddle (\textit{S} ) and the node (\textit{N} ) points for $\tau=0$ are also marked.}
\label{traj}
\end{figure}

\begin{figure}
\centerline{\includegraphics[width=12cm,height=8cm]{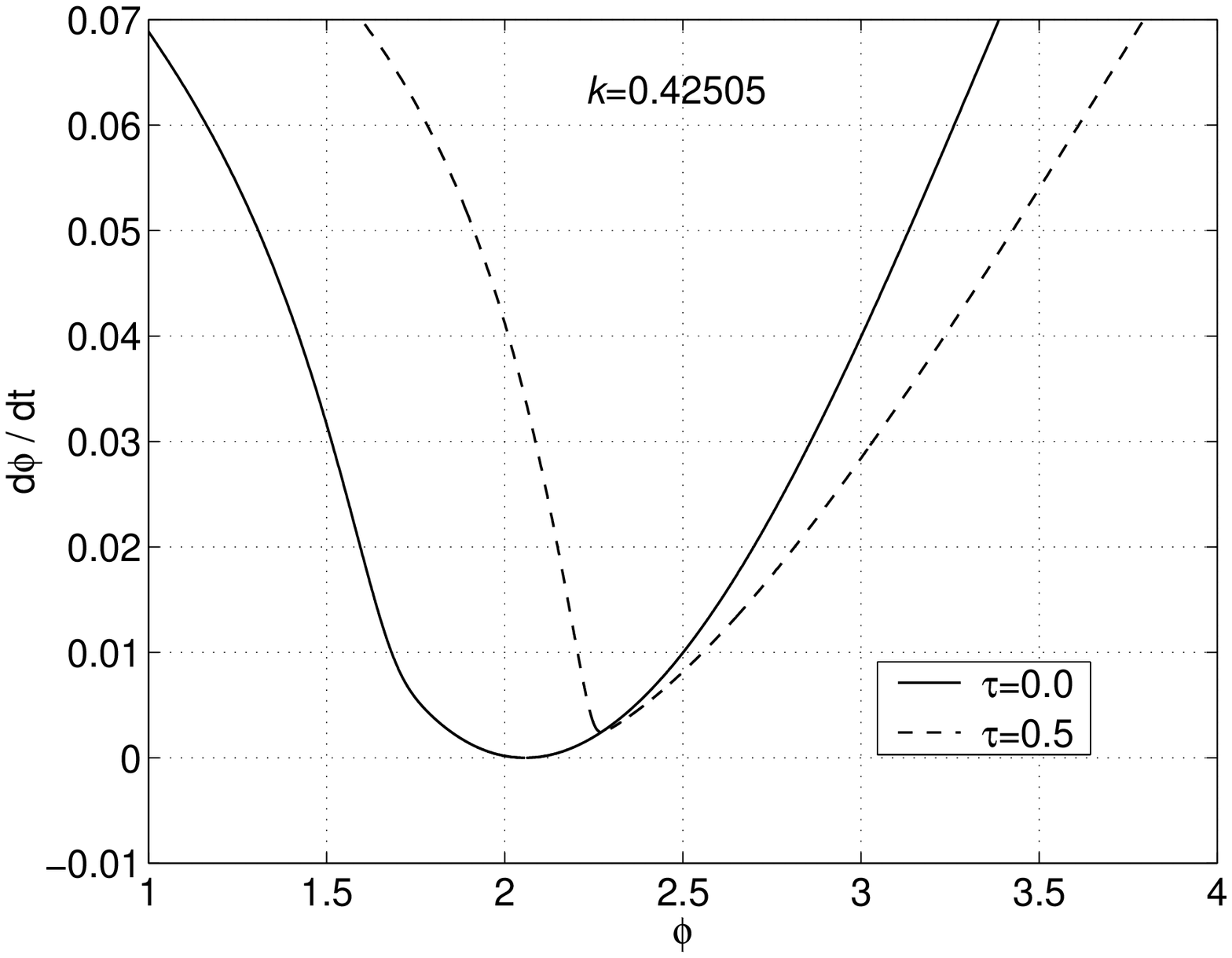}}
\caption{The rate of change of phase ($d\phi/dt$) as a function of phase($\phi$) near the bifurcation point.}
\label{freq_phi}
\end{figure}

\indent
The basic mathematical form of our neuron model is, 
\begin{equation}
\dot{z}(t)=\left[i(\omega+b\vert z(t)\vert ^{2})+\vert z(t)\vert ^{2}-\vert z(t)\vert ^{4}\right ]z(t)-kz^{2}(t-\tau) \label{feedback}
\end{equation} 
where $z=x+iy $ is the canonical fast variable representative of physical excitation quantities like the membrane potential, ionic channel currents etc., $\omega$ is the linear frequency of the oscillations and $b$ the shear parameter is a measure of the nonlinear frequency shift proportional to the amplitude of the oscillations. $k$ is the magnitude of the feedback strength and $\tau$ is a time delay parameter. In the absence of the feedback term the oscillator is poised at the subcritical Hopf bifurcation point and has the topological normal form for a Bautin bifurcation \cite{kuz}. 
In our model we have introduced a quadratic nonlinear feedback which is the simplest and lowest order nonlinearity that provides an excitable behavior. The feedback is time delayed to account for finite propagation times of signals. The self-feedback term can mock up a variety of physical effects. In a collection of coupled neurons, the term can be regarded as a source term representing the collective feedback due to the rest of the neurons \cite{ramana}. 
A more direct and natural application of our equation could be in the modeling of ${\it autapse}$ neurons where self-feedback can occur through auto-synapses.
The significance of self-feedback for ${\it autapse}$ neurons has recently been pointed out in \cite{chr2004}. To understand the fundamental dynamical basis of the spiking behavior we have first carried out a detailed bifurcation analysis of our model system in terms of the feedback strength parameter $k$ for different values of the time delay parameter $\tau$. Our results are shown in Fig. \ref{bif_k} where plots of the L2-norm ($\vert z \vert$)$\;{\it vs}\;$ $k$ are shown for $b=-0.5$ and $\tau=0, 0.5$ and $0.6$ respectively. In the absence of time delay ($\tau=0$)
we find that for $k>0.42506$ there exist two fixed points - a stable node and a saddle node. At $k=k_{c}=0.42506$ we have an SNLC bifurcation such that for $k<k_c$ one only has a stable limit cycle branch whose period tends to infinity at the bifurcation point. The location of $k_c$ depends on the value of the shear parameter $b$ but the nature of the bifurcation diagram remains the same as long as $-1<b \leq -0.5$.  The presence of finite time delay in the feedback introduces profound changes in the bifurcation diagram of the system. As seen from Fig. \ref{bif_k} the stable limit cycle branch now extends beyond the SNLC bifurcation point of $k=k_c$ and gives rise to a bistable region over this extension. The point $k=k_c$ now marks a saddle node (SN) bifurcation point. The range of extension increases as a function of $\tau$. 
The existence of the limit cycle in the region beyond $k=k_c$ has a threshold character in $\tau$ i.e. for a given $k$ beyond $k_c$ the limit cycle appears only when $\tau$ exceeds a critical value. 
The dynamical origin of this behavior can be traced to the onset of an SSL bifurcation \cite{izh2000} occurring as the parameter $\tau$ is varied. The two parameter bifurcation diagram in $k$ and $\tau$ space is shown in Fig. \ref{tau_k}. Different bifurcation branches as well as the regions of stable limit cycle/fixed point/bistable solutions are clearly demarcated.  Some typical trajectories for different values of $\tau$ and a fixed value of $k>k_c$ are shown in Fig. \ref{traj}. For $\tau$ values below a threshold all trajectories end up on the stable node and there is no oscillatory behavior. At a certain critical value of $\tau$ one notices the emergence of a separatrix loop trajectory and the consequent onset of periodic behavior. An analysis of the orbit dynamics close to the bifurcation point reveals another subtle time delay induced effect which has been noted before in \cite{ramana}. Writing $z=re^{i\phi}$, we plot the time rate of change of the phase ($d\phi/dt$) vs the variation in $\phi$ and display it in Fig. \ref{freq_phi}. In the absence of time delay the nature of the $d\phi/dt$ curve close to the bifurcation point is symmetric and nearly
parabolic. Finite time delay introduces a significant asymmetry as seen in the curve for $\tau=0.5$. This asymmetry, as we will discuss a litte later, influences the fine structure of the spike, namely, it introduces an asymmetry in the peak profile of an individual spike in terms of the rise and fall times (Fig. \ref{spikes}).
\indent

\begin{figure}
\centerline{\includegraphics[width=12cm,height=8cm]{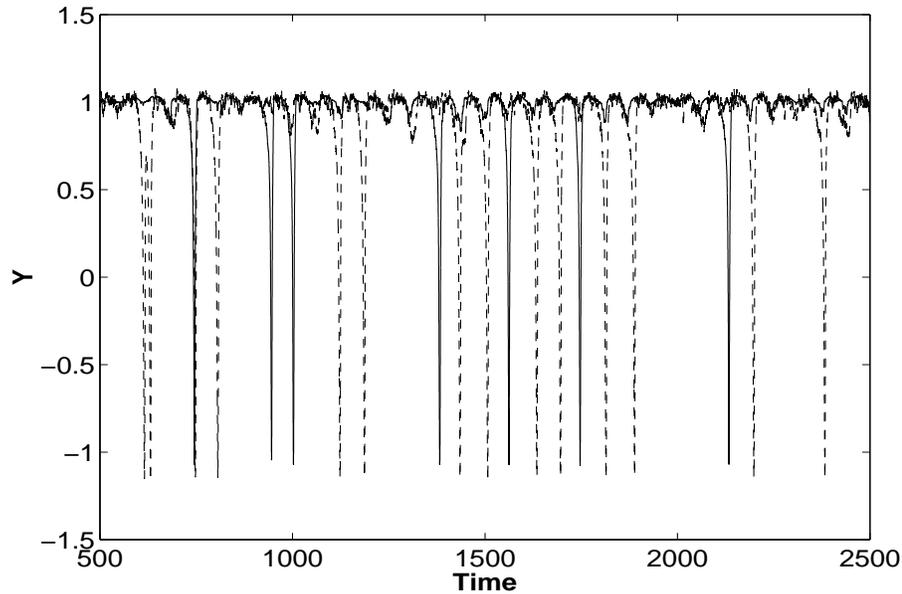}}
\caption{A segment of a typical time series exhibiting spiking. The continuous trace is for $\tau=0$ and the dashed trace is for $\tau=0.5$. Note the increase in the spiking rate in the presence of finite time delay in the feedback.}
\label{data}
\end{figure}

\begin{figure}
\centerline{\includegraphics[width=12cm,height=8cm]{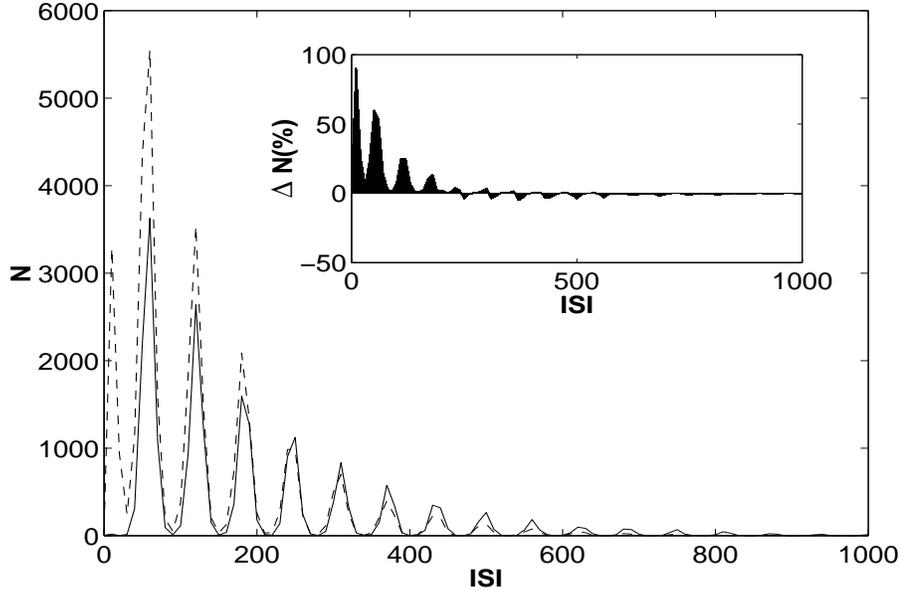}}
\caption{The interspike interval histograms (ISIHs) with delay(dashed curve) and without delay (solid curve).}
\label{hist}
\end{figure}

\begin{figure}
\centerline{\includegraphics[width=12cm,height=8cm]{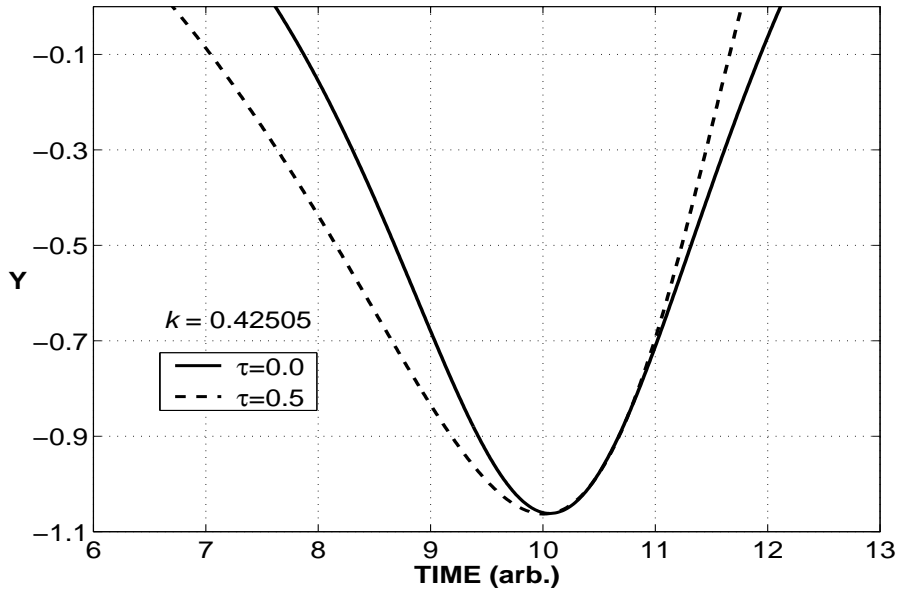}}
\caption{Peak profile of an individual spike in the absence of delay (solid curve) and in the presence of time delay (dashed curve).}
\label{spikes}
\end{figure}

We now discuss the dynamical consequences of these modifications in the bifurcation scenario on the neural excitability properties of the system. For this we introduce a source term $f(t)$ on the right hand side of eq.(\ref{feedback}) and study the temporal response of the neuron to this external stimulus. We choose $f(t)=\varepsilon e^{i\Omega t}+\sqrt{2D}\xi(t)$, where $\varepsilon$ is the amplitude and $\Omega$ is the frequency of a periodic signal and $\xi(t)$ is the zero mean Gaussian white noise with intensity $D$.
We drive the system at $k=0.45$ which is quite far from $k_c$ and choose the amplitude of the signal and the strength of the noise to be such that individually both are subthreshold and do not give rise to spiking on their own.  Spiking occurs under the combined influence of the external signal and noise in a manner akin to stochastic resonance phenomena \cite{longtin}. In other words the driver signal raises and lowers the potential of the system periodically and the noise helps in pushing it over randomly to the region $k<k_c$ where it executes a large periodic trajectory before returning to the rest state. A segment of a typical time series of this phenomenon is shown in Fig. \ref{data} 
where the amplitude of $y$ is plotted as a function of time for $\tau=0$ as a solid curve. The initial transients have been discarded for this plot and the parameter values used in the numerical integration of the model are $b=-0.5, \Omega=0.1 (T=2\pi/\Omega \simeq 62.8), \epsilon=0.04$ and $D=0.004$.
One clearly notices the spiking phenomenon which occurs near a preferred phase of the external stimulus and has a period of about $62.8$. A random number of periods are skipped between two successive firings. In the same figure we also show, by a dashed curve, the corresponding time series obtained for $\tau=0.5$. A marked difference between the two is the fewer number of periods skipped for the time delay case indicating an enhancement in the spiking activity. One also finds bursts of multiple spiking in the presence of finite time delay. For a more quantitative analysis we compute the ISIs of a large number of such time series ( about a 1000) by counting the times between two successive crossings of $y=0$ and construct a histogram (ISIH). Such a histogram, shown in Fig. \ref{hist}, exhibits a multimodal structure with peaks at integer multiples of the basic ISI period i.e. the period of the external stimulus which is about 62.8. The peaks of the ISIH are also seen to decay exponentially. A comparison of the $\tau=0$ and $\tau=0.5$ cases shows that the decay rate is faster for the latter. For the finite time delay case, the ISIs are also seen to be more concentrated at lower multiples of the basic ISI which is an indication of enhanced coherence in spiking with the external stimulus. A more quantitative measure of this delay induced change is displayed in the inset box of Fig. \ref{hist} where the percentage change over the first peak of no-delay case (corresponding to the basic ISI period, $T$, of the stimulus) is plotted and which show large deviations at the lower periods. In the histograms  we also notice an additional first peak for the finite time delay case and which is not located at an integer multiple of $T$. This peak arises due to the  multiple spikings with short ISIs that are seen in the time delayed traces of  Fig. \ref{data}. Similar signatures are also noticed in plots (not shown in the paper) of the return map of successive ISIs (with and without delay) where it is found that time delay in the feedback compresses the range of ISIs indicating a greater coherence. We should add here that the nature of the above results does not change with a change in the location of the working point away from $k=0.45$. If $k$ is chosen to be further away from $k_c$ then one needs to augment the amount of noise and/or the driver amplitude to observe the stochastic resonance behaviour of the spiking and similar time delay signatures are then obtained.
\indent

These delay induced changes in the spiking pattern can be understood in terms of the basic modifications in the bifurcation scenario. The extension of the limit cycle branch to the $k>k_c$ region for finite time delay can be seen as a lowering of the threshold for the stochastic resonance phenomena leading to a consequent statistical increase in the amount of spiking. In the absence of time delay or for low values of time delay (see Fig.2) the transition from rest state to limit cycle is through an SNLC bifurcation which for our driving parameters leads to a single spike. In the parameter regime where a bistable region exists the rest state goes to a limit cycle through an SN bifurcation and the limit cycle finally returns to the rest state through a SSL bifurcation thereby displaying a hysteretic behaviour \cite{frank}. The passage through the bistable region leads to multiple spiking
corresponding to the number of limit cycle periods spent on the return branch. The number of spikes can vary due to the effect of the external noise. In addition to these macroscopic statistical changes of the spiking behavior one also notices finer scale changes in the temporal structure of individual spikes. In Fig. \ref{spikes} a comparison of the peak profiles of an individual spike in the absence and presence of time delay shows a marked  asymmetry in the rise and fall times for the latter. This can be understood in terms of the modifications induced by time delay in the nature of the bifurcation and in the structure of the limit cycle orbit. The dynamical reason for the sharp temporal shape of a spike is the highly nonuniform motion of the system along a limit cycle orbit \cite{rappel}, namely, the system spends most of its time near the bifurcation point and then quickly traverses the rest of the angular region. The rate of approach and departure from this `bottleneck' region is quite asymmetric for the time delayed case, as seen in Fig. \ref{freq_phi}, and this manifests itself in the pronounced asymmetry of the temporal shape of the spike.
\indent

To summarize, we have investigated the effect of time delay on the excitability properties of a mathematical model consisting of a subcritical Hopf oscillator with a nonlinear time delayed feedback. We find that time delay can have a significant influence on the spiking properties of the system, such as in enhancing the frequency of spikes, triggering of multi-spikes and altering the fine structure of individual potential spikes all of which can have interesting practical implications in real biological systems as well as in modeling studies of ensembles of artificial neurons. Our model neuron could also provide a useful paradigm for gaining more insight into the behavior of autapse neurons which are presently receiving a great deal of theoretical and experimental attention.
\indent

We are grateful to B. Ermentrout and T. Luzyanina for their help on the use of XPPAUT  \cite{xppaut} and DDE-BIFTOOL \cite{dde} respectively.

\end{document}